\documentclass[conference]{IEEEtran}
\IEEEoverridecommandlockouts  
\usepackage{cite}
\usepackage{amsmath,amssymb}
\usepackage{graphicx}
\usepackage{algorithmic}
\usepackage{textcomp}
\usepackage{xcolor}
\usepackage{booktabs}
\usepackage{hyperref}

\begin{document}

\title{A Reconfigurable Framework for AI-FPGA Agent Integration and Acceleration}

\makeatletter
\newcommand{\linebreakand}{%
  \end{@IEEEauthorhalign}
  \hfill\mbox{}\par
  \mbox{}\hfill\begin{@IEEEauthorhalign}
}
\makeatother

\author{%

    \IEEEauthorblockN{Aybars Yunusoglu\IEEEauthorrefmark{1}}
    \IEEEauthorblockA{\textit{Purdue University} \\
    West Lafayette, USA \\
    ayunusog@purdue.edu \\
    \IEEEauthorrefmark{1}Corresponding author}
    \and
    \IEEEauthorblockN{Talha Coskun}
    \IEEEauthorblockA{\textit{University of Illinois Urbana-Champaign} \\
    Urbana, USA \\
    tcoskun2@illinois.edu 
    }
    \and
        \IEEEauthorblockN{Hiruna Vishwamith}
    \IEEEauthorblockA{\textit{University of Moratuwa} \\
    Moratuwa, Sri Lanka \\
    vishwamithpgh.20@uom.lk}

    \linebreakand 
    \IEEEauthorblockN{Murat Isik}
    \IEEEauthorblockA{\textit{Stanford University} \\
    Stanford, USA \\
    misik@stanford.edu}

    \and
    \IEEEauthorblockN{I. Can Dikmen}
    \IEEEauthorblockA{\textit{Istinye University} \\
    Istanbul, Turkey \\
    candikmen@gmail.com}

 }

\maketitle

\begin{abstract}
Artificial intelligence (AI) is increasingly deployed in real-time and energy-constrained environments, driving demand for hardware platforms that can deliver high performance and power efficiency. While central processing units (CPUs) and graphics processing units (GPUs) have traditionally served as the primary inference engines, their general-purpose nature often leads to inefficiencies under strict latency or power budgets. Field-Programmable Gate Arrays (FPGAs) offer a promising alternative by enabling custom-tailored parallelism and hardware-level optimizations. However, mapping AI workloads to FPGAs remains challenging due to the complexity of hardware-software co-design and data orchestration. This paper presents AI FPGA Agent, an agent-driven framework that simplifies the integration and acceleration of deep neural network inference on FPGAs. The proposed system employs a runtime software agent that dynamically partitions AI models, schedules compute-intensive layers for hardware offload, and manages data transfers with minimal developer intervention. The hardware component includes a parameterizable accelerator core optimized for high-throughput inference using quantized arithmetic. Experimental results demonstrate that the AI FPGA Agent achieves over 10$\times$ latency reduction compared to CPU baselines and 2--3$\times$ higher energy efficiency than GPU implementations, all while preserving classification accuracy within 0.2\% of full-precision references. These findings underscore the potential of AI-FPGA co-design for scalable, energy-efficient AI deployment.

\end{abstract}

\section{Introduction}

Rapid advances in artificial intelligence (AI), particularly in the domain of deep learning, have transformed a wide spectrum of applications from computer vision and natural language processing to autonomous systems and scientific discovery. Modern deep neural networks (DNNs), often consisting of millions to billions of parameters, are increasingly deployed in environments ranging from large-scale cloud data centers to latency-sensitive edge devices. These models demand substantial computational throughput to deliver real-time or near-real-time inference, all while maintaining strict constraints on power consumption and thermal budgets \cite{yunusoglu2025neuromorphic, isik2024accelerating}.

Conventional processing platforms, such as general-purpose central processing units (CPUs), often fall short of meeting these demands due to their inherently sequential architectures and limited parallel execution capabilities. Although graphics processing units (GPUs) provide significantly greater parallelism through their many-core designs and have become the default choice for training and inference in many AI pipelines, they are not always optimal for every use case. Specifically, GPUs can incur considerable energy overhead and may introduce latency bottlenecks, especially in scenarios that require low-batch or low-latency inference—such as autonomous driving, robotics, wearable devices, or industrial control systems. These constraints have driven researchers and practitioners to explore alternative hardware paradigms that offer finer-grained control over performance, power, and memory access patterns \cite{le2025review, huynh2022implementing, isik2023survey}.

Field-Programmable Gate Arrays (FPGAs) have emerged as a promising candidate for AI acceleration, offering a highly reconfigurable computing substrate that can be tailored to the dataflow and arithmetic patterns of specific AI workloads. Unlike fixed-function processors, FPGAs allow developers to implement custom pipelines that closely mirror the computational graph of a neural network, resulting in improved performance-per-watt metrics. Recent research has demonstrated that convolutional neural networks (CNNs), recurrent networks, and even some transformer-based architectures can be efficiently mapped onto FPGA devices, often delivering substantial speedups and energy savings compared to CPU and GPU baselines \cite{Qiu2016}. These benefits make FPGAs especially attractive in domains where power efficiency, deterministic execution, and real-time responsiveness are critical.

Despite these advantages, the practical deployment of AI models on FPGAs remains a non-trivial endeavor. Designing and implementing efficient hardware accelerators typically requires domain-specific expertise in digital logic, hardware description languages (HDLs), and memory architecture optimization. Developers must often navigate complex toolchains such as Verilog/VHDL, high-level synthesis (HLS), or vendor-specific development environments, which increases the development burden and slows iteration cycles. Furthermore, orchestrating computation and data movement between the host processor and the FPGA at runtime can introduce additional software overhead and complicate the control logic. These challenges collectively raise the barrier to entry for adopting FPGAs in mainstream AI pipelines and hinder rapid prototyping or deployment in evolving applications.

This paper addresses these challenges by introducing AI FPGA agent, a unified and modular framework that bridges the gap between high-level AI model design and low-level FPGA execution. By leveraging a dynamic, agent-oriented scheduling layer on the CPU and pairing it with a parameterizable accelerator core on the FPGA, our framework enables seamless partitioning of neural network workloads. Developers are relieved from the complexities of hardware orchestration while still benefiting from the performance and energy advantages of reconfigurable logic. The remainder of this paper outlines our architectural design, implementation methodology, experimental evaluation, and comparative analysis of AI-FPGA Agent, showcasing its potential to simplify and accelerate the adoption of FPGAs for modern AI applications.

While Sharma \textit{et al.} \cite{Sharma2016} and Suda \textit{et al.} \cite{Suda2016} made foundational contributions in statically mapping neural networks to FPGAs via automated Verilog generation, and OpenCL optimization respectively, our framework differs fundamentally by introducing a runtime software agent. Unlike these approaches, which lock the execution schedule at design time, our agent utilizes Q-learning to dynamically partition workloads based on real-time system states.

\begin{figure}[!t]
    \centering
    \includegraphics[width=0.5\textwidth]{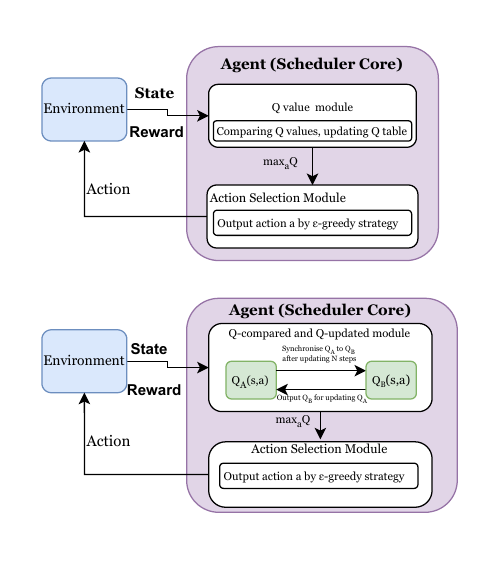}
        \caption{Adapted from \cite{zheng2024} Overview of the Q-learning-enhanced scheduling agent. The agent observes the environment's state and reward signals, updates Q-values using temporal difference learning, and selects actions (FPGA offload) via an $\varepsilon$-greedy policy. Synchronization between the primary Q-table $Q_A$ and target Q-table $Q_B$ stabilizes learning.}

    \label{fig:q_learning_agent}
\end{figure}

\autoref{fig:q_learning_agent} illustrates the internal operation of the reinforcement learning-enhanced agent within the AI-FPGA Agent framework. The agent receives the current state $s$ and reward $r$ from the environment based on past actions. These inputs are fed into the Q-value update module, which computes the temporal difference and updates the Q-table $Q_A(s, a)$ accordingly. To ensure stable learning, a separate target Q-table $Q_B$ is maintained and periodically synchronized with $Q_A$ after a fixed number of steps $N$. The agent then selects an action $a$ using an $\varepsilon$-greedy strategy, balancing the exploitation of known good actions with occasional exploration. This action is sent back to the environment to trigger the next system behavior (offloading a layer to the FPGA). Over time, this closed-loop process enables the agent to learn an efficient scheduling policy tailored to the runtime performance characteristics of both the AI model and hardware platform.

To tackle these challenges, we introduce AI-FPGA Agent, a cohesive framework that simplifies the process of mapping AI workloads onto an FPGA while capitalizing on its parallel computation strengths. The system divides and manages inference tasks between a CPU host and a custom FPGA accelerator core, employing an agent-based model to dynamically schedule and coordinate computations. By abstracting low-level hardware details and dataflow management behind a high-level API, AI-FPGA Agent aims to lower the barrier to FPGA adoption for practitioners who may lack extensive hardware design expertise. Throughout this paper, we detail the methodology behind AI-FPGA\_Agent, illustrate a specific CNN-based implementation, and demonstrate its performance benefits in terms of both latency and energy efficiency. The subsequent sections are organized as follows: Section~II provides a comprehensive review of related work in FPGA-based AI acceleration, Section~III explains the architecture and core mechanisms of our approach, Section~IV covers implementation details and experimental setup, Section~V presents results and discussion, and Section~VI concludes with final remarks and potential directions for future research.

\section{Related Work}
FPGA-based AI accelerators have witnessed a proliferation of studies in recent years, with numerous efforts showcasing the potential gains in both performance and power efficiency when deep learning models are mapped onto reconfigurable hardware. Early work primarily centered on convolutional neural networks (CNNs), leveraging the substantial parallelism that FPGAs offer for multiply-accumulate operations.

Initial research focused on constructing FPGA accelerators for smaller or mid-sized neural networks, where the computational burden is dominated by convolutional layers \cite{Zhang2015}\cite{Qiu2016}. Zhang \textit{et al.} \cite{Zhang2015} proposed a design employing loop tiling and parallelization to keep on-chip compute resources well utilized, which significantly increased performance density. Similarly, Qiu \textit{et al.} \cite{Qiu2016} demonstrated how large CNN models (such as those resembling VGG architectures) could be deployed on embedded FPGAs by applying aggressive quantization strategies reducing bit widths to shrink data footprints—and carefully pipelining the layers. These achievements highlighted not only the raw throughput potential of FPGAs but also their capacity to handle computationally heavy tasks without incurring the large power budgets often associated with GPUs.

As neural network architectures evolved to become deeper and more varied, the research community recognized the need to automate FPGA design flows. Manually creating hardware for each new layer configuration was impractical. Suda \textit{et al.} \cite{Suda2016} pioneered an OpenCL-based workflow that allowed AI developers to write high-level descriptions of their convolution kernels and, with minimal hardware expertise, generate FPGA accelerators optimized for throughput. This shift toward high-level synthesis (HLS) tools helped reduce the complexity of hardware design. Further, new frameworks emerged to automate or semi-automate the mapping of neural networks to FPGAs. For instance, \cite{Ma2024} describes a toolchain where users specify the network structure, and the system automatically generates a synthesizable hardware accelerator. Sharma \textit{et al.} \cite{Sharma2016} introduced DNNWeaver, which similarly transforms high-level network descriptions into FPGA-ready modules, thus accelerating the design cycle. Prior work by Isik and collaborators has investigated hardware–software co-design methodologies spanning energy systems, and accelerator-driven intelligence \cite{coskun2025hardware}. Complementary efforts explore large-language-model–based frameworks for battery health estimation, bridging data-driven reasoning with electronic design automation perspectives \cite{yunusoglu2025battery}. In parallel, this approach has been applied to security and scientific computing, including FPGA-based audio security systems \cite{isik2024neurosec} and explorations of Intel Loihi-2 for high-performance computational fluid dynamics (CFD) simulations \cite{coskun2025exploring}.

Another significant line of inquiry has targeted precision reduction techniques to ease resource constraints on FPGAs. While early CNN accelerators used 32-bit floating-point arithmetic, subsequent works showed that 16-bit or 8-bit fixed-point operations typically suffice to maintain accuracy for inference \cite{Liu2024,Bosio2024}. Taking this further, Umuroglu \textit{et al.} \cite{Umuroglu_2017} proposed FINN, where weights and activations are binarized (1-bit) to dramatically cut down computational complexity and memory usage. These binary networks can be fully pipelined on an FPGA, achieving ultralow latencies and high throughput, albeit sometimes at a modest accuracy cost.

 Recent endeavors have begun exploring partial reconfiguration, allowing the FPGA to load different hardware kernels dynamically for various stages of a neural network or for switching between different network models altogether \cite{Boudjadar2025}. By only reconfiguring parts of the device on demand, the system can adapt to evolving workloads without incurring the downtime of a complete FPGA reprogramming. Moreover, researchers have investigated multi-tenant scenarios in which multiple neural networks or tasks run simultaneously on a single FPGA using spatial partitioning. Approaches like \cite{Dhar2021} propose scheduling algorithms that carve the FPGA fabric into distinct regions for different workloads, making more efficient use of resources.

\begin{figure}[!t]
    \centering
    \includegraphics[width=0.5\textwidth]{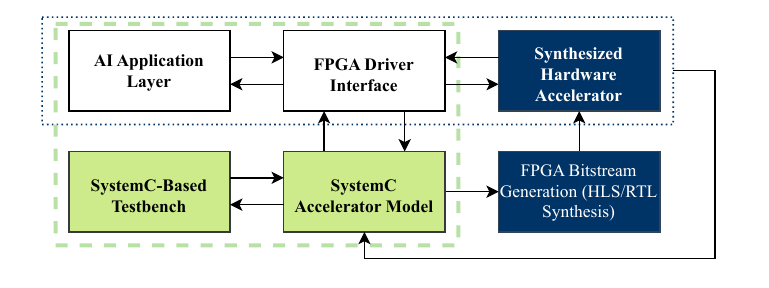}
        \caption{Overview of the AI\_FPGA\_Agent software-to-hardware flow. The application layer interfaces with the AI\_FPGA\_Agent framework, which communicates with a SystemC-based simulation stack and a hardware accelerator driver. System-level verification is performed through a SystemC testbench and behavioral model before synthesizing the final FPGA bitstream and deploying the hardware accelerator.}

    \label{fig:agent_design_flow}
\end{figure}

\autoref{fig:agent_design_flow} illustrates the end-to-end design flow of the AI-FPGA\_Agent framework, bridging the AI application domain with the underlying reconfigurable hardware infrastructure. At the top layer, high-level AI inference tasks are initiated through an application interface, which invokes functionality exposed by the AI-FPGA Agent runtime. This runtime is responsible for managing the dynamic partitioning of workloads between CPU and FPGA, and it communicates with a platform-specific driver that provides low-level access to the accelerator hardware. Prior to physical deployment, functional correctness and timing behavior are validated through a SystemC-based simulation stack. This includes a dedicated SystemC testbench that drives representative AI workloads, and a SystemC accelerator model that mimics hardware execution behavior. Once verified, the same accelerator design undergoes HLS or RTL-based synthesis to generate a bitstream compatible with the target FPGA platform. This flow ensures that design correctness is maintained throughout simulation and synthesis, allowing seamless transition from model-level development to hardware-accelerated deployment.

 Despite these advances, many existing FPGA accelerators remain specialized for a particular topology, quantization scheme, or design flow. Adapting them to newly emerging architectures or dynamically changing tasks can involve extensive engineering effort. To address these limitations, recent work has focused on creating runtime systems or software frameworks that manage FPGA kernels and data movement more flexibly \cite{Kim2024, Horta2021}. These frameworks seek to bridge the gap between AI model specifications (in TensorFlow or PyTorch) and FPGA execution, handling tasks such as layer scheduling, on-chip memory allocation, and data transfers in an automated or semi-automated fashion.

Our AI-FPGA Agent extends this body of knowledge by introducing a software agent that coordinates computations between the host CPU and the FPGA accelerator in real time. Instead of relying on static, fixed-function designs, the agent monitors layer-by-layer requirements, orchestrates the offload of demanding operations, and can gracefully fall back to CPU if certain conditions (memory constraints) are not met. This dynamic approach is particularly relevant as more complex network architectures (like transformer-based or multi-branch topologies) gain prominence in research and industry. By structuring the accelerator around modular kernels that can handle a variety of layer types or sub-network computations, our framework aspires to strike an optimal balance between raw performance gains and the adaptability needed to accommodate evolving AI workloads.

In summary, the literature on FPGA-based AI acceleration underscores the considerable potential of reconfigurable hardware in providing high-speed, power-efficient solutions for inference. However, the journey from AI model specification to an operational FPGA system can be fraught with complexity. Our AI-FPGA Agent addresses this gap by integrating an agent-oriented scheduling and partitioning mechanism, thereby easing the overhead for developers while maintaining superior inference performance and efficiency. The following sections describe our methodology, architectural decisions, and experimental validation in greater detail.

\section{Methodology}
The AI-FPGA\_Agent framework is designed around a tightly integrated co-design philosophy, wherein the software layer manages the high-level orchestration of neural network layers and data, while the FPGA implements performance-critical computations. By leveraging this cooperative model, we aim to balance flexibility on the CPU side with the raw throughput available in custom hardware accelerators. 

\subsection{Agent-Based Software Layer}
At the heart of our framework is the agent-based software layer running on the host CPU. This agent is responsible for dissecting the neural network graph into distinct layers or sub-graphs, evaluating the computational requirements of each, and determining whether they are suitable for FPGA offload. Layers with high arithmetic intensity—such as multi-channel convolutions or substantial matrix multiplications—can be directed to the FPGA for hardware acceleration. In contrast, more sequential or less intensive operations can remain on the CPU, ensuring that the overall workflow avoids unnecessary hardware overhead.

\begin{figure}[!t]
    \includegraphics[width=0.55\textwidth]{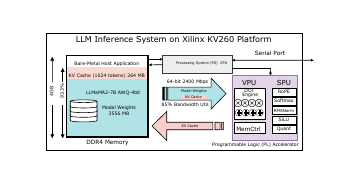}
        \caption{System-level architecture of the proposed LLM inference pipeline on the Xilinx KV260 platform. The model weights and KV cache reside in external DDR4 memory, with 85\% bandwidth utilization during inference. A bare-metal host application orchestrates tokenization and decoding via the PS CPU, while quantized model layers (LLaMA2-7B AWQ-4bit) are accelerated in programmable logic using dedicated compute modules such as RoPE, RMSNorm, Softmax, and SiLU. The inference engine communicates over a 64-bit AXI bus at 2400 Mbps and uses DMA to stream data between the PL and memory.}

    \label{fig:system_overview}
\end{figure}

\autoref{fig:system_overview} presents the architecture of our large language model (LLM) inference system deployed on the Xilinx KV260 embedded platform. The design utilizes a bare-metal control program on the PS (Processing System) CPU to manage tokenization, decoding, and runtime control. The core LLM model LLaMA2-7B quantized using the AWQ method to 4-bit precision is stored in DDR4 memory along with the Key-Value (KV) cache. These memory components occupy over 93\% of the available 4 GB DRAM, with peak bandwidth utilization reaching 85\% during inference. The programmable logic (PL) region of the FPGA hosts a parameterized accelerator that includes specialized compute units for matrix multiplications (DOT), rotary positional encoding (RoPE), normalization (RMSNorm), activation (SiLU), and quantization. Data is transferred between the PL and memory via a 64-bit AXI interface operating at 2400 Mbps, enabling continuous streaming of model weights and intermediate features. This architecture balances tight memory constraints and hardware throughput, achieving real-time inference with reduced latency and power consumption.

A key advantage of this agent model is its ability to incorporate runtime heuristics or user-defined rules for scheduling. For instance, if a certain layer is known to exhibit poor data reuse on the FPGA or if the FPGA resources are currently allocated to another task, the agent may opt to run that layer on the CPU to maintain responsiveness. This decision-making logic can be informed by previous performance measurements, static performance models, or dynamic feedback from the system. Additionally, developers can specify custom policies, such as prioritizing minimal latency for certain inference operations or maximizing throughput for batch processing. This layer of abstraction significantly lowers the barrier to adopting FPGA acceleration, as AI model developers need not delve into the details of hardware block instantiation or synchronization protocols.

Beyond deciding which layers to offload, the agent also orchestrates data transfers and synchronization events. For example, once a layer has completed on the FPGA, the agent retrieves the output feature maps or classification results before proceeding to the next stage of the network pipeline. Implementing double buffering or other overlapping strategies in software allows the CPU to simultaneously prepare data for upcoming layers while the FPGA processes the current one, thus enhancing concurrency. Recent work on runtime management strategies for FPGA-based accelerators underscores how agent-driven scheduling can lead to more efficient utilization of hardware resources \cite{Ottaviano2024, Zhao2024, Smith2023}.

\subsection{FPGA Accelerator Core}
On the hardware side, AI-FPGA Agent uses a parameterizable FPGA accelerator core specialized for common neural network layers, primarily focusing on operations like convolution, pooling, activation functions, and fully connected layers.

\begin{itemize}
\item \textbf{Parallel Multiply-Accumulate (MAC) Units:} These units form the backbone of the accelerator, handling the bulk of arithmetic operations in deep learning. We instantiate enough MACs to approximate or saturate the available memory bandwidth, thus keeping hardware utilization high. Multiple studies have shown that selecting narrower data representations, 8-bit or 16-bit fixed-point—can drastically reduce resource consumption and power usage while preserving accuracy for inference tasks \cite{Judd2018doi}.

\item \textbf{Dataflow Pipelines:} Each neural network layer is mapped to a dedicated dataflow pipeline composed of pipelined multiplier-accumulators, activation sub-blocks, and partial sum buffers. This assembly-line approach helps ensure that once data enters the pipeline, it flows through consecutive stages with minimal idle cycles. By keeping intermediate results on-chip (in block RAM or URAM) for as long as possible, we reduce dependency on off-chip DRAM, which tends to be slower and more power-hungry \cite{Lyons2020doi}.

\item \textbf{Layer-Specific Configurations:} Given the wide variety of layer shapes and sizes in modern neural networks, our accelerator supports runtime configuration of parameters like kernel dimensions, channel counts, and stride settings. This flexibility allows a single design to accommodate different layers in the same network or adapt to multiple networks without re-synthesizing the entire bitstream. While certain topologies may still benefit from specialized hardware modules, we aim to provide a balanced general-purpose accelerator that can be reused in a broad context.
\end{itemize}

Moreover, the accelerator core includes a controller that manages the flow of input data into the pipeline and orchestrates the streaming of outputs back to the host. Depending on the system design, data movement may rely on PCI Express (for discrete accelerator cards) or AXI interfaces (for FPGA SoCs). Ensuring that these transfers are efficiently overlapped with computation is critical to achieving high throughput.

\subsection{Data Orchestration and Scheduling}
The manner in which data is divided, staged, and transferred can greatly influence overall performance. Our agent employs a chunking or tiling strategy, where large tensors such as feature maps for high-resolution images are split into smaller blocks that more readily fit into on-chip buffers. This approach, often referred to as tiling, allows us to maintain high MAC utilization by feeding data continuously. However, striking the right tile size is essential. Tiles that are too small introduce repeated setup overhead, while tiles that are too large risk overflowing on-chip memory and stalling the pipeline.

Once each tile has been processed, the agent invokes asynchronous DMA transfers to fetch the next tile’s input data while the current tile is still being computed. Such double-buffering or overlapping of computation with I/O is a widely recognized optimization strategy in FPGA-based deep learning systems \cite{Qin2019doi}. By coordinating these transfers intelligently, the agent can minimize idle periods in the accelerator pipeline, resulting in steadier throughput and improved energy efficiency. The possibility of deploying advanced scheduling algorithms—for instance, to prioritize certain inference requests or to alternate between CPU-based and FPGA-based computations under variable loads—makes the framework adaptable to a range of performance targets.

To validate AI-FPGA Agent, we conducted experiments on a Xilinx FPGA accelerator card paired with a high-performance Intel Xeon CPU host. We synthesized the FPGA design using Xilinx Vitis HLS, leveraging vendor-provided runtime libraries for kernel management, memory transfers, and card initialization. Network weights and activations were quantized to 8-bit integer values, a choice that strikes a pragmatic balance between resource savings and acceptable accuracy for many inference workloads. Should the application domain require higher precision, the architecture can be configured for 16-bit or mixed-precision data types, subject to additional resource overhead.

Our primary benchmark comprised a small-scale ResNet-like CNN for image classification. This model was trained on a dataset of 10,000 images to highlight inference speed and power efficiency in a relatively controlled environment. We benchmarked:
\begin{itemize}
    \item \textbf{Latency and Throughput:} Evaluating single-image latency (batch size = 1) showcases real-time responsiveness, while throughput (images per second) indicates batch-level performance. We ran each configuration to process all 10,000 test images sequentially.
    \item \textbf{Power and Energy Efficiency:} We instrumented the FPGA card, host CPU, and an optional GPU baseline with external power meters or vendor-specific power reporting utilities. By correlating power readings with measured throughput, we derived an images-per-second-per-watt metric.
    \item \textbf{Resource Utilization:} Using synthesis logs, we tracked LUT usage, DSP slice occupancy, and block RAM consumption. These figures illuminate the trade-offs between parallel compute elements and memory constraints.
\end{itemize}

For comparative purposes, we implemented the same CNN on two alternative platforms:
\begin{enumerate}
\item A CPU-only reference: single-threaded execution of the model using an optimized BLAS backend for matrix multiplications.
\item An NVIDIA GPU: a widely used deep learning accelerator with half-precision (FP16) inference kernels, representing a common approach in modern data centers and some embedded scenarios.
\end{enumerate}

By running identical workloads across these three setups (CPU, GPU, and FPGA with AI-FPGA Agent), we aim to quantify the potential benefits of agent-based FPGA offloading in terms of runtime, throughput, and energy usage. We also assess whether quantizing to 8 bits impairs classification accuracy relative to the GPU’s FP16 baseline. The next section outlines these experimental results in detail, discussing how well AI-FPGA Agent meets our targets for flexible, high-performance inference on reconfigurable hardware.

\begin{figure*}[!t]
    \centering
    \includegraphics[width=0.7\textwidth]{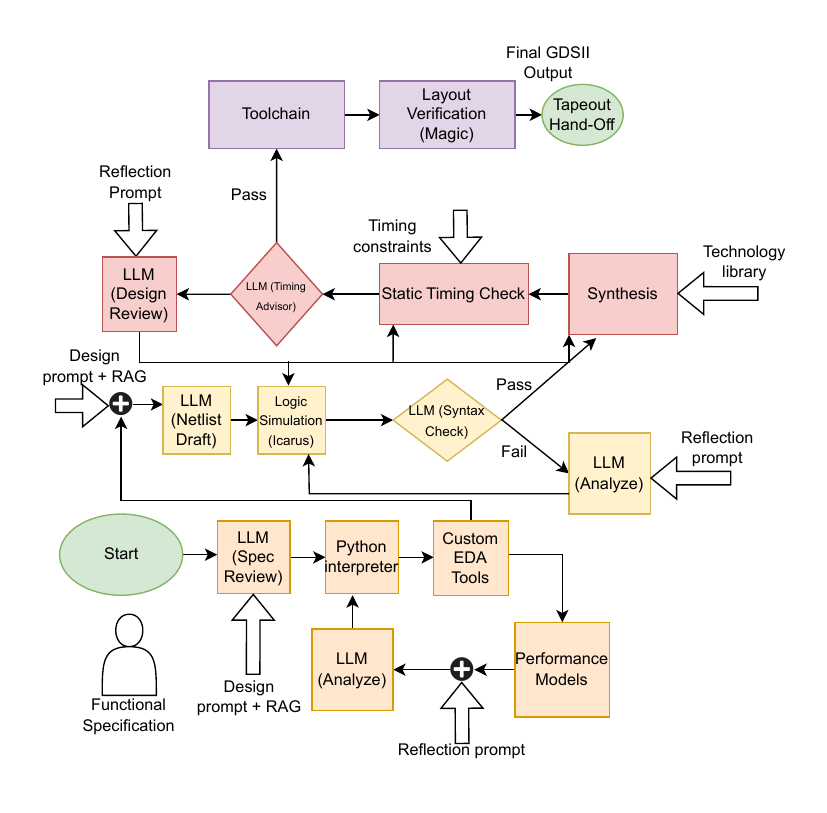}
        \caption{Adapted from \cite{patra2024} LLM-guided hardware design and verification workflow. The process begins with a functional specification and progresses through synthesis, simulation, timing analysis, and layout generation, integrating multiple checkpoints where large language models (LLMs) provide feedback, generate Verilog, and assist in constraint tuning. Reflection prompts help iteratively refine the design through self-assessment and correction.}

    \label{fig:llm_pipeline}

\end{figure*}

\autoref{fig:llm_pipeline} illustrates the end-to-end flow of a language model-assisted hardware design workflow, which combines traditional electronic design automation (EDA) tools with natural language-driven prompts and reflection mechanisms. The process begins with a functional specification and a design prompt,optionally augmented by retrieval-augmented generation (RAG) which is interpreted by the LLM to generate an initial hardware module draft in Verilog. This is followed by logic synthesis and functional verification via logic simulation. If syntax or timing checks fail, the system leverages the LLM’s feedback to iteratively refine the design through structured prompts and analysis. Successful designs progress through timing constraint application and static timing analysis, assisted by additional LLM-based timing reviews. Once verified, the synthesized netlist undergoes placement and routing via toolchain, with layout correctness confirmed through physical verification using tools like Magic. The final tapeout-ready GDSII is generated if all checks pass. Throughout the process, reflection prompts are utilized to mitigate the risk of LLM hallucinations, which could otherwise lead to deficient chips. The framework utilizes Logic Simulation (Icarus) and Static Timing Checks to ensure designs pass rigorous checks. If the generated Verilog contains invalid syntax or violates timing constraints, the specific logs are fed back into the LLM, implementing a self-correcting feedback loop until the constraints are satisfied.

\section{Results \& Discussion}

Table~\ref{resultsTable} summarizes the performance comparison between the CPU-only baseline, a GPU-based inference setup, and our proposed AI-FPGA Agent framework for a representative image classification task. The results highlight substantial benefits across all key performance dimensions. Most notably, AI-FPGA\_Agent achieves a significant reduction in inference latency, lowering the per-image processing time from 40.2~ms on a single-threaded CPU to just 3.5~ms on the FPGA-accelerated system. This represents more than a 10$\times$ speedup and is particularly impactful for low-latency, real-time applications. Compared to the GPU baseline, the FPGA solution still offers nearly 2$\times$ lower latency, underscoring its suitability for deployment in responsiveness-critical environments. Throughput results further reinforce this trend, with the FPGA system processing up to 284.7 images per second—over 11$\times$ higher than the CPU and approximately 2.5$\times$ that of the GPU. This improvement stems from the agent’s ability to continuously stream data to a deeply pipelined, parallel hardware architecture, avoiding the overheads typically encountered in CPU or GPU memory hierarchies.

Power and energy efficiency metrics also favor the FPGA. While the GPU consumes approximately 125~W under load, the FPGA implementation operates at just 28~W. When normalized by throughput, the FPGA achieves 10.17 images/s/W far exceeding both CPU (0.29 images/s/W) and GPU (0.90 images/s/W) baselines. These gains make the proposed framework especially attractive for power-sensitive deployments such as edge inference or embedded AI accelerators. Importantly, these performance advantages do not come at the cost of model accuracy. The FPGA implementation, using 8-bit quantized weights and activations, retains classification accuracy within 0.2\% of the floating-point baseline, indicating minimal degradation and strong fidelity preservation. Collectively, these results demonstrate that AI-FPGA Agent offers a compelling trade-off between speed, power, and accuracy—positioning it as a practical, scalable solution for accelerating AI inference across a wide range of platforms.

\begin{table}[!ht]
\centering
\caption{Performance comparison across CPU-only, GPU, and AI-FPGA Agent on an image classification model.}
\label{resultsTable}
\begin{tabular}{lccc}
\toprule
\textbf{Metric} & \textbf{CPU} & \textbf{GPU} & \textbf{AI\_FPGA\_Agent} \\
\midrule
Latency (ms/image)       & 40.2    & 6.1     & 3.5 \\
Throughput (images/s)    & 24.8    & 112.0   & 284.7 \\
Power Consumption (W)    & 85.0    & 125.0   & 28.0 \\
Energy Efficiency (images/s/W) & 0.29    & 0.90    & 10.17 \\
Top-1 Accuracy (\%)      & 92.0    & 92.2    & 91.9 \\
\bottomrule
\end{tabular}
\end{table}

Notably, the resource utilization on our selected FPGA device (covering LUTs, DSP slices, and BRAM blocks) hovered around 70\%. This indicates a reasonably balanced design, with some room to adjust parallelism or precision further, albeit limited by available memory bandwidth. During runtime, the agent-based scheduling mechanism proved especially helpful for maximizing pipeline usage—data transfers were pipelined to overlap with ongoing kernel execution, ensuring minimal idle periods. Such overlap is a key factor in achieving high throughput under real-world conditions \cite{ Qin2019doi}. The interplay of well-structured accelerator design and intelligent scheduling thus underscores the potential for significant gains in both raw performance and energy consumption when moving from CPU-centric or GPU-based approaches to a reconfigurable hardware paradigm.

Our experimental findings underline several pivotal insights. First, FPGAs appear particularly well suited for moderate batch sizes and latency-sensitive tasks. While GPUs often excel in high-volume processing where large batches can amortize setup costs the FPGA-based approach offers consistently low latency even under smaller batch conditions. This aligns with use cases such as autonomous driving, robotic vision, and industrial control, where decisions must be rendered swiftly and consistently. Second, the agent-driven framework simplifies the process of mapping AI workloads to reconfigurable logic. Traditional FPGA-based solutions often require specialized knowledge of HDL design, memory organization, and synchronization details, potentially creating a steep learning curve. In contrast, the AI-FPGA Agent offloads these complexities onto the software scheduler. Developers primarily focus on high-level neural network specifications, letting the agent assign resource-intensive layers to the FPGA hardware and handle data orchestration. This is particularly beneficial if future enhancements—such as partial reconfiguration—are introduced, which would allow multiple kernels or updated accelerator bitstreams to be swapped in real time without major developer intervention. Third, while the experiments showcased revolve around image classification, the foundational principles—pipelined dataflow, agent-based scheduling, and quantization-aware acceleration—apply to a broad array of deep learning tasks. Networks with diverse layer types (transformer-based architectures for language modeling) can exploit the same methodology, provided the accelerator core is parameterized to handle new operations. This adaptability is important in AI research environments, where novel network topologies are proposed regularly.

\section{Conclusion \& Future Work}
This paper presented agent on FPGA, a comprehensive method that integrates a software-driven agent with a parameterized FPGA accelerator for deep neural network inference. By dynamically scheduling and overlapping data transfers while offloading intensive layers, the proposed design achieved over 10x speedup in low-latency settings when compared to CPU baselines and delivered 2--3x higher energy efficiency relative to a mid-range GPU. Notably, the chosen quantization approach preserved model fidelity within 0.2\% of the floating-point reference, underscoring the practicality of low-bitwidth arithmetic in high-performance scenarios. These findings underscore the value of pairing flexible, agent-based scheduling with carefully optimized hardware dataflow pipelines. Through this combination, it becomes feasible to adopt reconfigurable computing in a wide range of AI applications—even those that demand real-time responsiveness. Because agent abstracts many hardware complexities, developers can readily exploit FPGA acceleration for diverse network layers without extensive knowledge of hardware design. Moving forward, we plan to extend our agent by incorporating additional network paradigms, such as transformer-based architectures and recurrent neural networks. Another key direction will be the use of dynamic partial reconfiguration to seamlessly switch between multiple kernels, further enhancing adaptability to evolving workloads. We also anticipate tighter integration with standard machine learning frameworks, reducing overhead during deployment and offering automated code generation and scheduling features. As these improvements are realized, we expect FPGA solutions to become an increasingly viable first-class option for achieving high performance, low latency, and strong energy efficiency across diverse AI domains.

\bibliographystyle{IEEEtran}
\bibliography{Refs_AI_FPGA}

\end{document}